%% file: main.tex
\documentclass[10pt,conference]{IEEEtran}
\IEEEoverridecommandlockouts
\usepackage[letterpaper, left=0.625in, right=0.625in, bottom=1in, top=0.75in]{geometry}

\usepackage{cite}
\usepackage{amsmath,amssymb,amsfonts}
\usepackage{algorithm2e}
\usepackage{graphicx}
\usepackage{textcomp}
\usepackage{verbatim}
\usepackage{makecell}
\usepackage{graphicx}
\usepackage{subcaption}
\usepackage{xcolor}
\usepackage{multirow}
\usepackage{array}
\usepackage{color, colortbl}
\usepackage{verbatim}
\definecolor{Gray}{gray}{0.9}
\usepackage{alphalph}
\usepackage{multicol}
\usepackage{amsmath}
\usepackage{balance}
\usepackage{hyperref}

\RestyleAlgo{ruled}

\def\BibTeX{{\rm B\kern-.05em{\sc i\kern-.025em b}\kern-.08em
    T\kern-.1667em\lower.7ex\hbox{E}\kern-.125emX}}

\usepackage{amsmath}
\usepackage{amsfonts}

\begin{document}

\title{Active Learning Framework to Automate Network Traffic Classification}

\author{
\IEEEauthorblockN{Jaroslav Pešek \quad Dominik Soukup}
\IEEEauthorblockA{
\textit{CTU in Prague} \\
Thakurova 9, Prague, Czech Republic \\
pesekja8@fit.cvut.cz \quad soukudom@fit.cvut.cz}

\and
\IEEEauthorblockN{Tomáš Čejka}
\IEEEauthorblockA{
\textit{CESNET} \\
Zikova 4 Prague, Czech Republic \\
cejkat@cesnet.cz}
}

\maketitle

\begin{abstract}
Recent network traffic classification methods benefit from machine learning (ML) technology. However, there are many challenges due to use of ML, such as: lack of high-quality annotated datasets, data-drifts and other effects causing aging of datasets and ML models, high volumes of network traffic etc. This paper argues that it is necessary to augment traditional workflows of ML training\&deployment and adapt Active Learning concept on network traffic analysis. The paper presents a novel Active Learning Framework (ALF) to address this topic. ALF provides prepared software components that can be used to deploy an active learning loop and maintain an ALF instance that continuously evolves a dataset and ML model automatically. The resulting solution is deployable for IP flow-based analysis of high-speed (100\,Gb/s) networks, and also supports research experiments on different strategies and methods for annotation, evaluation, dataset optimization, etc. Finally, the paper lists some research challenges that emerge from the first experiments with ALF in practice.

\end{abstract}


\IEEEpeerreviewmaketitle

\section{Introduction}
\label{sec:introduction}
\input{Intro.tex}

\section{Related Work}
\label{sec:related}
\input{Related.tex}

\section{Active Learning Framework (ALF)}
\label{sec:alf}
\input{ALF.tex}

\section{Targeted Real Use-Cases}
\label{sec:usecases}
\input{useCase.tex}

\section{Real World Deployment}
\label{sec:deployment}
\input{realDeploy.tex}

\section{Open Research Challenges}
\label{sec:future}
\input{future.tex}

\section{Conclusion}
\label{sec:conclusion}
Active learning principle has been known to the academic community for many years. However, its application in network traffic analysis appeared quite recently. Network monitoring and traffic analysis (e.g., for network security purposes) are crucial areas that can benefit from machine learning technology. Moreover, active learning approach allows for continuous updates of both datasets and machine learning models.

Therefore, we have developed a new Active Learning Framework (ALF) to address the needs related to machine learning application in the network traffic analysis domain. ALF is designed to process a stream of extended flow data (commonly used in practice to monitor large network infrastructures) and to evolve datasets and machine learning models using an annotator, optimization methods, and real online traffic. The reason for ALF use instead of some alone annotator lies in the performance limits, and computational or time complexity of the annotating (i.e., labeling process), and finally possibility to retrieve ground truth information that is not available for all data in practice. Therefore, ALF is meant to train machine learning models using available information by the annotator so that such models can be deployed in the other network environments where no annotator can be used. 

The paper also lists several research areas related to the active learning in network traffic and, more specifically, to ALF. We believe the current version of ALF will help to accelerate research activities in areas like quality of dataset assessment, research of replacement strategies and dataset optimization, and so on. 

\section*{Acknowledgement}

This work was supported by the European Union’s Horizon 2020 research and innovation program under grant agreement No.~833418 and also by the Grant Agency of the CTU in Prague, grant No. SGS20/210/OHK3/3T/18 funded by the MEYS of the Czech Republic.

\ifCLASSOPTIONcaptionsoff
  \newpage
\fi

\bibliographystyle{IEEEtran}
\balance
\bibliography{_biblio}

\end{document}

%% file: Intro.tex
Machine learning (ML) is modern technology, and worldwide researchers show its feasibility in plenty domains including network traffic monitoring and analysis. Especially, the popularity of end-to-end encryption in network traffic forces us to leave traditional methods of traffic analysis based on visibility into Application layer (L7) (as it was feasible in the past, e.g., in \cite{NEMEA}) or exploiting unencrypted information from connection establishment (e.g., TLS handshakes in \cite{ETA}). The decryption of data is not feasible in practice on scale, and machine learning is a promising way to exploit advanced statistics derived from encrypted communication flows even on high speed communication links (presented, e.g., in \cite{eta-ml}).

On the other hand, ML is dependent on the target domain and dataset that is used for training. In most cases, the dataset must be annotated for network traffic classification (for supervised ML). This requirement makes ML usage very complicated since we need to provide ground truth labels. Without reliably annotated datasets, it is not possible to gain relevant results from ML models.

It is common that the research experiments for scientific papers are performed using a fixed dataset, i.e., from one fixed (even short) period of time, which works perfectly in laboratory environments. However, network traffic evolves, as it is studied by Brabec et al. in \cite{Brabec2020}, and we can intuitively argue that ML models and datasets are becoming obsolete in time. Also, deployment in a different network can cause a performance decrease. As a result, even a perfectly annotated dataset with good performance will drop in accuracy due to aging. Therefore, we recognize an essential requirement to research methods for autonomous ML deployment, monitoring, and regular updates over time.

Fortunately, there is a promising concept of Active Learning (AL) as a sub-field of ML to deal with a huge amount of incoming data. The aim of AL is to reduce the initial need of labeled data records by intelligent querying labels during the training phase (AL is described in more details in Sec.~\ref{sec:related}). This approach can continuously add new data instances and build a relevant up-to-date dataset for any target domain.  

AL is defined as a general concept that leaves a wide space for methods and possibilities of how entities of AL are used (annotator, query strategy, ML model, and input data). Many state-of-the-art approaches are incomplete or unavailable for our network traffic analysis use case; thus, some implemented system is missing for practical deployment in production. Therefore, we propose an Active Learning Framework (ALF)\footnote{\url{https://github.com/CESNET/ALF}} focused primarily on network traffic analysis and designed for stream-based processing that allows online or offline dataset evaluation. 



This paper describes the following main contributions:
\begin{itemize}
     \item We propose Active Learning Framework prototype --- novel flexible and modular framework for network traffic classification:
        \begin{itemize}
            \item based on Active Learning technology and state-of-the-art principles and methods;
            \item allowing for autonomous and continuous dataset creation and evolving;
            \item supporting automatic updates, evaluation, and optimization of annotated datasets of network traffic;
            \item supporting extensive monitoring of ML models and all stages in ALF workflow.
        \end{itemize}
     \item Using ALF, we evaluate different ``Query strategies'' during dataset update, and present new findings contrary to related work. 
     \item We propose more types of annotators to standardize annotation process within ALF.
     \item We showcase experiments on real data and use cases to prove feasibility of ALF. 
\end{itemize}

This paper is a broad extension of our previous work
\cite{Soukup2021}, which provided initials ideas for quality of datasets and ML model deployment.

The paper is organized as follows. Section~\ref{sec:related} lists related works in active learning, quality of datasets, and related solutions. Section~\ref{sec:alf} contains details about the architecture and main features of the ALF system. Section~\ref{sec:usecases} lists the most common use cases in the networking domain we target and evaluation of ALF. Section~\ref{sec:deployment} presents results we have achieved in a real network environment in online mode. Finally, Section~\ref{sec:future} mentions related open research challenges, and Section~\ref{sec:conclusion} concludes the paper.

%% file: Related.tex
This section describes related papers to AL methods but also datasets quality since our framework is focused on more advanced use cases. 

\subsection{Active Learning (AL)}

AL is a generally known concept in ML domain. Settles \cite{Settles2009ActiveL} provided the first comprehensive survey in this field. This survey includes available methods and scenarios that can be applied with AL. The author discusses a \textit{pool-based}, resp. \textit{stream-based} approaches to AL, which are designed for processing an offline batch of data, resp. online continuous stream of data. Even though pool-based approaches are commonly chosen by researchers, the author recommended stream-based approaches as more appropriate for cases like ours. He also described basic AL \textit{Query strategies} (i.e., methods to select which data to annotate), and three of them are relevant for stream-based use cases (uncertainty sampling, information density, and random sampling).

Recently, a new technical study regarding AL-based methods for network traffic classification has been published by Shahraki et al. \cite{AlStudy}. The authors reviewed several Query strategies, and empirically tested and evaluated them. Based on this study, we extend our set of query strategies with the Reinforcement AL (RAL) method.

Ju el al. \cite{datadrift}, used AL method together with concept drift. Wasserman et al. \cite{wassermannRALImprovingStreamBased2019} introduced a concept reinforcement active learning (RAL) which utilizes a feedback from an annotator. Cardoso et al. \cite{cardosoRankedBatchmodeActive2017} defines ranked batch for removing duplicates (in sense of distance similarity) in batch of selected flows in one iteration.

\subsection{Quality of datasets}
Quality of datasets is important area for real deployment of ML classifiers. After initial definitions from Soukup et al. \cite{Soukup2021}, Camacho et al. \cite{Camacho2022} proposed a permutation based network dataset quality assessment with promising results. Li et al. \cite{Li} proposed a distance-entropy method to distinguish the good and bad data. This method is described as introductory and calls out for further investigation in this domain.
Dataset quality evaluation in combination with AL is beneficial to address challenge of stopping criteria to retrain ML classifier and verify if dataset is better. 

Many researchers are focused on data cleaning and dataset optimization that is relevant extension of AL approach. Yoon et al. \cite{DVRL} proposes a novel meta learning framework for data evaluation that is jointly optimized with the
target task predictor model. The proposed framework is focused on removing noisy and incorrect labels. 
Our framework is based on trusted annotator. However, it could be beneficial to use these methods in optimization section of our framework described in Sec. \ref{sec:alf}. 

Several papers are separately focused  on dataset cleaning which typically involves dealing with data inconsistencies, duplicates, outliers or missing values, but also class imbalance \cite{Sahu2020}, class overlapping \cite{Dudjak2021}, incorrect labeling \cite{Cordeiro2020}, noise in data \cite{Gupta2019}, unnecessary instances \cite{instanceSelection}, and dataset size problem \cite{Sun2017}. These challenges are analyzed separately and general framework that will allow to implement them together is missing. There are existing AL libraries such as \textit{Scikit-learn}\footnote{https://scikit-learn.org/stable/}, \textit{modAL}\footnote{https://github.com/modAL-python/modAL} or \textit{ALiPy}\footnote{https://github.com/NUAA-AL/ALiPy}. However, none of them is flexible enough to implement methods to address challenges with building dataset. Muller et al. \cite{muller} analyzed five steps (Discovery, Capture, Curation, Design, Creation) that are done to create datasets. This workflow was investigated with 21 data science
professionals. The dataset creation is very individual task and based on author's reputation. In this work we propose our framework to standardize building of dataset and their monitoring with optimization after deployment.

Ginart et el. \cite{MLmonitoring} introduced the need for post-deployment ML monitoring. This work proposes a new method with theoretical guarantees to detect changes in input data stream and ML results, such as distribution or data drift. ML model monitoring is part of our framework too. Moreover, it is connected with AL. Therefore, it leverages information from annotators and human experts to automatically improve the dataset based on detected changes. 

\subsection{Summary of related work and differences from our research}

ML and AL domains are very rich and popular among researchers. Our proposed solution is based on the available state of the art ideas. According to \cite{Settles2009ActiveL}, we focused primarily the stream-based approach to minimize required capacity of storage for high speed network traffic. The other mentioned related works were taken as inspiration for design and development of the working prototype based on AL.

We delivered a universal and flexible framework, ALF, that implements known AL query strategies suitable for stream-based AL and combines state-of-the-art knowledge referenced in the previous sections. We implemented strategies described by Settles and Shahraki et al. Based on our experiments, better results are obtained with a batch of flows. For informative selection in batch, we implemented Cardoso's ranked batch. We also implemented Wassermann's RAL with a referential implementation with non-significant changes. Principles introduced by Wassermann's RAL might be used in the future with more strategies. The AL core is extended by optimization and monitoring modules to allow post-deployment improvements of the dataset and ML model.

Using ALF, we can start with an initial dataset and a simple ML model and automatically improve it over time with higher visibility. Also, ALF was tested on real traffic --- IP flow data from multiple high-speed links (order of 100\,Gb/s) at the perimeter of ISP network, and outcomes are compared with results from the current state of the art. 
The proposed ALF in this paper extends the works \cite{Settles2009ActiveL,AlStudy} and targets several open challenges. Details are provided in Sec.~\ref{sec:alf}.

To our best knowledge, there is no publicly available software implementation of active learning technology applied and optimized for i) network traffic classification, ii) continuous evolving annotated network traffic datasets for ML, iii) support of research in dataset evaluation and optimization. Such ambitions were set for the design and development of the proposed ALF described in our paper.

%% file: ALF.tex
In this work, we propose a framework that aims to improve the deployment and monitoring of ML models in real networks. This section provides more detailed description of its design and components. 
\subsection{Overview}
\label{sec:overview}

\begin{figure}[h]
\centering
  \includegraphics[width=3.0in]{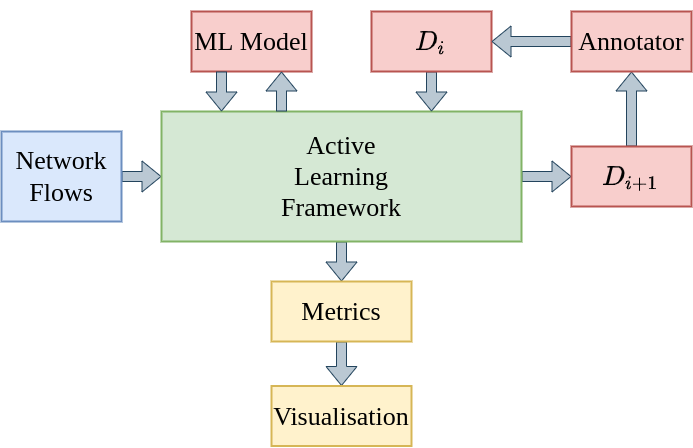}%
  \caption{Example of Active Learning Framework workflow}
  \label{fig:al-oveview}
\end{figure}

ALF is designed as a modular framework that can be completely customized. Generally, it provides AL workflow that works with any input data format and ML model. However, this paper is mostly focused on the network traffic domain. An example of the workflow is depicted in high-level scheme in Fig.~\ref{fig:al-oveview}.

\begin{figure}[h]
    \centering
    \includegraphics[width=8cm]{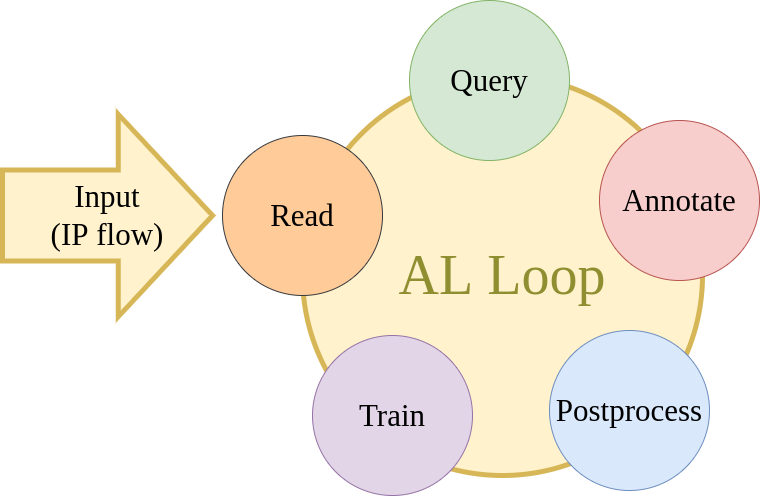}
    \caption{Active Learning Loop (AL loop) implemented in ALF}
    \label{fig:alloop}
\end{figure}

As an input, we expect an initial dataset ($D_0$) that can be even empty, annotator, target ML classifier, and unlabeled data. Based on the provided configuration, the framework can work in offline (pool-based) or online (stream-based) mode. This brings flexibility since we can process network data directly from the network interface or load stored traffic in pcap format. More details are described in Sec.~\ref{sec:monitoring}. The input is standardized to feature the dataset using a preprocessing engine, i.e., the data must be converted to the supported data format (e.g., CSV) before processing in ALF. Once the preprocessing is done, ALF starts an \textit{AL loop} shown in Fig.~\ref{fig:alloop}.

AL loop consists of the following \textbf{stages} that are internally performed repeatedly within ALF:
\begin{enumerate}
    \item \textit{Read} receives new IP flow record representing network communication; this record will be classified using ML model in ALF.
    \item \textit{Query} selects (samples) flow records that should be annotated and maybe included into the dataset (according to a chosen \textit{Query strategy}, described later).
    \item \textit{Annotate} provides ground truth for the currently selected flow record.
    \item \textit{Postprocess} aims to assess quality of the current dataset and optimize it after possible insertion of newly annotated flow record.
    \item \textit{Train} prepares new version of ML model $M_{i+1}$ trained with the current dataset $D_{i+1}$.
\end{enumerate}

It is worth noting that the AL Loop may skip some phases. For example, evaluation and optimization tasks (within the Postprocess phase) are too time consuming, and it is preferable to run them regularly only once per long time period (e.g., in hours, days, or even weeks) based on a Query strategy and score function. This loop generates new versions of the dataset $D_{i+1}$, computes metrics for monitoring, and performs dataset optimization and quality assessment.

Dataset quality assessment is based on permutation testing introduced by Camacho et al. \cite{Camacho2022}. It allows to evaluate the quality of each dataset version and compare the improvement for possible retraining of the classifier. Standard AL approach is focused on enhancing an existing dataset with new valuable data. Nevertheless, this would incrementally increase the size of the dataset, its redundancy and complexity. Therefore in the proposed ALF, we include an option for dataset optimization methods that remove unnecessary records and try to find the minimal dataset as described by Soukup et al. \cite{Soukup2021}.

The process of obtaining ground  truth information is very individual among data scientists (as it is stated by \cite{muller}). The understanding of dataset quality usually depends more on author's reputation than on data and labels. Contrary, we argue that the quality of the dataset should be assessed using statistical and structural properties of the content of the dataset that can be evaluated and by careful monitoring the performance of the deployed ML model in production.

ALF uses the concept of annotators to standardize the process of labeling. Generally, an annotator is a software module that is able to annotate a subset of input data (network flows in our case) based on some additional information from external sources (e.g., OSINT, system logs, service logs, or any monitoring/auditing tools running at end-point devices). Naturally, having a perfect annotator seems like an ideal solution to the whole classification task. However, in practice, the annotator is not available for all cases and cannot be deployed to process the whole traffic. Therefore, it is assumed annotator is not able to provide ground truth to all data, and some \textit{Query strategy} is required to select the subset that can be and should be annotated. For example, an annotator based on information from end-points cannot be used in the network environment with devices out of control. The use of decryption to annotate the traffic is not feasible at a large scale.
Our idea is to train the ML model using an annotator in some controlled environment and, afterward, deploy and update classification models in the production. This approach allows automated and consistent labeling that AL methods can leverage.

Collected metrics from all stages are used for monitoring of classifier and dataset performance over time. These metrics include Matthews correlation coefficient, $\text{F}_1$ score, accuracy, recall, and precision. These values are visualized for different query strategies over time to compare.

Even though our solution is focused on automation and monitoring of all steps, there is still a possibility of mislabeling data and getting false positive (FP) detection results. In such a case, we allowed an external feedback loop from human operators in ALF. This feature triggers remarking of labels for specific data records according to provided knowledge. For example, this feedback is useful for security teams of a Security Operations Center (SOC) that manually investigate reported security incidents and are able to identify false alerts that ML reported from ALF.  


From the technical point of the overview, ALF has been developed in Python language and was primarily intended for any UNIX-based operation system. The whole framework is implemented as a single Python module with several classes. These classes are cross-referenced and exchange data via class attributes. Based on this approach, ALF can be used as a standalone module but can also be integrated with other frameworks, e.g., NEMEA\footnote{NEMEA system is an open source IP flow based tool for stream-wise analysis of network traffic.} \cite{NEMEA} is a working example.


\subsection{Architecture}
\label{sec:architecture}

AL loop, described in Sec.~\ref{sec:overview}, has to be supported by ALF and prepared for users/developers. Therefore, there are existing components that provide default functionality and leave the possibility to customize the stages of the AL loop for specific use cases. The architectural blocks are explained in this section.

\begin{figure}[h]
\centering
  \includegraphics[width=3.0in]{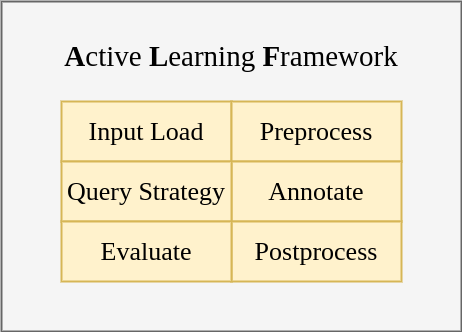}%
  \caption{High level architecture of ALF components }
  \label{fig:al-architecture}
\end{figure}

Figure \ref{fig:al-architecture} shows the simplified architecture of ALF with the main modules that are described further. As of now, one instance of ALF is meant for one ML classifier, more specifically, one classification task in network traffic. However, the ALF instance can duplicate stages, e.g., to run more Query strategies simultaneously within the instance. Additionally, the whole framework can be easily deployed in parallel for more ML classifiers. ALF is designed as a modular framework that can be customized for each deployment.  

\subsection*{Components for Read stage}
The \textit{Input Load} component is responsible for loading input data into a standardized format for further processing. Current implementation supports 
loading online stream of data from the NEMEA system or offline data from the filesystem.
Offline loading can be done from \textit{CSV} or \textit{UniRec} files (binary format of NEMEA).

Loaded data is typically processed to create a \textit{feature dataset} (defined in \cite{Soukup2021}). This operation is performed in the \textit{Preprocess} component. It includes custom methods for each ML use case to convert input data (e.g., network flows) to feature dataset. Data scientists usually develop these methods during a prototyping stage. 

\subsection*{Component for Query stage}
The \textit{Query Strategy} component is part of the AL core of this framework. It contains an implementation of AL strategies to select the most valuable records for annotation and update of the dataset. We implemented the most popular query strategies based on the state-of-the-art, such as Uncertainty sampling, Query-By-Committee, Random, Information Density, and Reinforcement AL (RAL) (the strategies were explained in \cite{Settles2009ActiveL, wassermannRALImprovingStreamBased2019}). Evaluation and performance results of these methods are described in Sec.~\ref{sec:usecases}. Preferably, all the implemented strategies are stream-based to minimize large data storage requirements that would otherwise be needed for the pool-based processing. 

\subsection*{Components for Annotate stage}
An important part of ALF is an automatic mechanism to annotate input data and evolve a reliable dataset. This task is handled by the \textit{Annotate} component. In this work, we deployed two types of annotators. The first is based on information from external services. For example, to annotate DoH traffic, we leverage a database of known DoH resolvers that can be further verified with some active checks (scans). This type of annotator can be a direct part of ALF, and the service (database) can be hosted anywhere. The second type of annotator is hosted on managed endpoints where we can gather a lot of information about network traffic before it is encrypted. For example, we developed a browser web extension plugin\footnote{https://github.com/jan-kala/WebTrafficAnnotator} that can collect and analyze all requests created by the web browser. The collected data is paired with a local IP Flow exporter, and the result is sent to the IP Flow collector, which provides data to ALF. Using the second type, we continuously annotate traffic even before the \textit{Read} stage of the AL loop.

\subsection*{Components for Postprocess stage}
After selecting new data records and modifying (updating) the dataset, we need to verify whether the update helped or not. This is the purpose of the \textit{Evaluate} component. In the current version of ALF, the evaluation is based on a metric score from the confusion matrix; however, we plan to include the dataset quality assessment methods natively in the future to have higher visibility and confidence in decisions about updates.

Based on the provided results from the \textit{Evaluate} component, we can further process and optimize the dataset in module \textit{Postprocess}. During the postprocessing, we store generated metrics (Matthews Correlation Coefficient, F1 Score, accuracy, recall, and precision) and logs for long-term monitoring of ALF instances performance. Also, when we add new data instances, we need to control dataset size and imbalance ratio. Currently, we use standard methods of oversampling and undersampling. 



\subsection{Monitoring and Configuration}
\label{sec:monitoring}
One of the main contributions of this work is a flexible framework that supports extensive measurement and monitoring of the ML models performance metrics and datasets properties. The configuration supports to set required methods and processing based on the target use case. Each module described in Sec. \ref{sec:architecture} has its own class attributes, which are implemented by all classes in a particular module. The user defines these parameters before each experiment.

Performance monitoring is a native part of ALF. During all stages, we collect metrics, as described in Sec. \ref{sec:architecture}. These metrics are persistently stored into a local database together with a trained ML classifier ($\text{M}_i$). With this approach, ALF can be restarted without affecting/loosing already processed data. Valuable metrics are visualized using Grafana\footnote{https://github.com/grafana/grafana} dashboard. 



%% file: useCase.tex
ALF already contains several algorithms that are available in each component to be chosen for specific needs of a classification problem. This section describes use cases we have targeted and that helped to evaluate the proposed framework and its features.

\subsection{Classification Improvement (even from empty dataset)}
\label{sec:improvement}
 ALF instance is able to run even in such configuration with empty or insufficiently small initial dataset. Annotator with a suitable query strategy then works as a generator of a completely new initial dataset inside ALF. This way, it is possible to generate annotated datasets automatically from the real network traffic, incrementally update them and to evolve machine learning models.

For demonstration of this use case, we used ALF for classification of DoH traffic against non-DoH HTTPS communication. First, we captured DoH data to discover an appropriate query strategies and their parameters. Experimenting in offline mode on captured data could be considered as white-box testing. We have discovered several surprising observations:
\begin{itemize}
    \item approaches based on similarity (such as ranked batch and information density) do not work well; parameter $\beta$ (see Settles \cite{Settles2009ActiveL}) has actually negative effect,
    \item threshold based on score has no significant effect which is in contrary to Settles; thus, future research might focus on suitable alternatives,
    \item random query strategy is fast and works sufficiently although is easily surpassed by uncertainty based strategies (least confident scoring, entropy scoring, RAL, etc.)
\end{itemize}


\subsection{ML Model Development}
\label{sec:devel}
\begin{figure}
    \centering
        \includegraphics[width=0.45\textwidth]{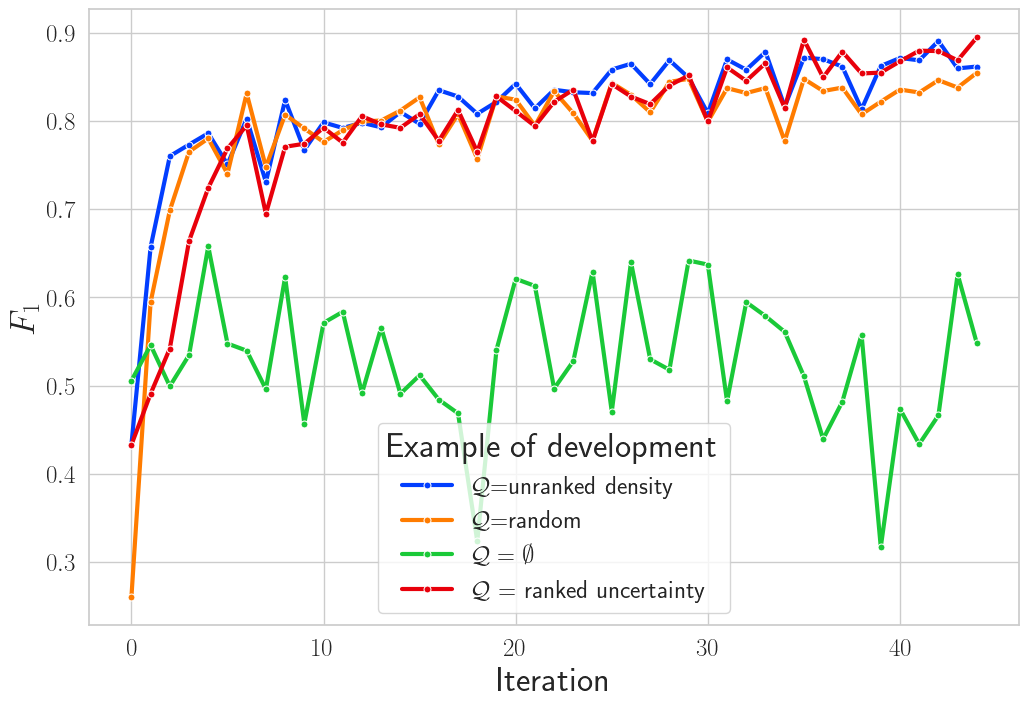}
    \caption{Example of development of a model. We compared different Query strategies (and run without any strategy used --- $\emptyset$) on the same data.}
    \label{fig:doh-development}
\end{figure}

It is generally complicated to build a precise ML model and keep it updated for long time. ALF can be used as annotator and build an ML model quickly with low effort since we assume query strategy selects most informative samples. Advantage of this approach is saving effort with annotating flows -- in conventional approach we assume all flows to be annotated. In this use case we do not strictly demand stream-based AL. Even though ALF is developed with consideration of streams, employing pool-based AL is also possible. In Fig.~\ref{fig:doh-development}, we demonstrate a model development through iterations. 

\subsection{Offline Testing}
\label{sec:testing}
ALF may work in offline mode which is useful for tuning parameters of prediction models and/or sample strategies. We collected and classified traffic of cryptomining protocol. Cryptominer dataset then consists of balanced amount of labels. Experiments were executed repeatedly, results are shown in Tab.~\ref{tab:cryptominers}.

\begin{table}[]
\caption{Comparison of Query strategies during the offline experiments with Cryptominer detection. All measurements were executed on the same dataset and the same iteration size = 10k which trigger AL loop}
\label{tab:cryptominers}
\centering
\begin{tabular}{ccc}
\hline
Strategy             & Average final $F_1$  & Average query time [$sec/it$]\\ \hline
unranked uncertainty & 0.952  & 0.001          \\
ranked uncertainty   & 0.919  & 2.116         \\
unranked density     & 0.699  & 2.592         \\
ranked density       & 0.642  & 3.069         \\
KL divergence        & 0.919  & 2.820        \\
random               & 0.860  & 0.001         \\ \hline
\end{tabular}
\end{table}

%% file: realDeploy.tex
ALF framework has already been experimentally deployed for pilot testing on a flow collector in CESNET2\footnote{CESNET2 is a national research and education network infrastructure in the Czech Republic} network infrastructure. The network has many network lines in range from 10\,Gb/s to 100\,Gb/s
connecting CESNET2 to other backbone networks. On average, the monitoring probes at the perimeter of the network generate
about 300,000 IP flow records per second during peak hours.
This pilot deployment improved debugging and revealed new requirements and feature requests.

The deployed ALF instance is related to the DNS over HTTPS (DoH) research of our team. Having a list of known DoH resolvers (i.e., providers of DoH service) and also using an active scan, it is possible to check whether a connection is expected to be a DoH. This process of checking creates an annotator required for ALF. This way, there is a training of a machine learning model that is constantly evolving to detect DoH communication. 

\begin{figure}
    \centering
        \includegraphics[width=0.45\textwidth]{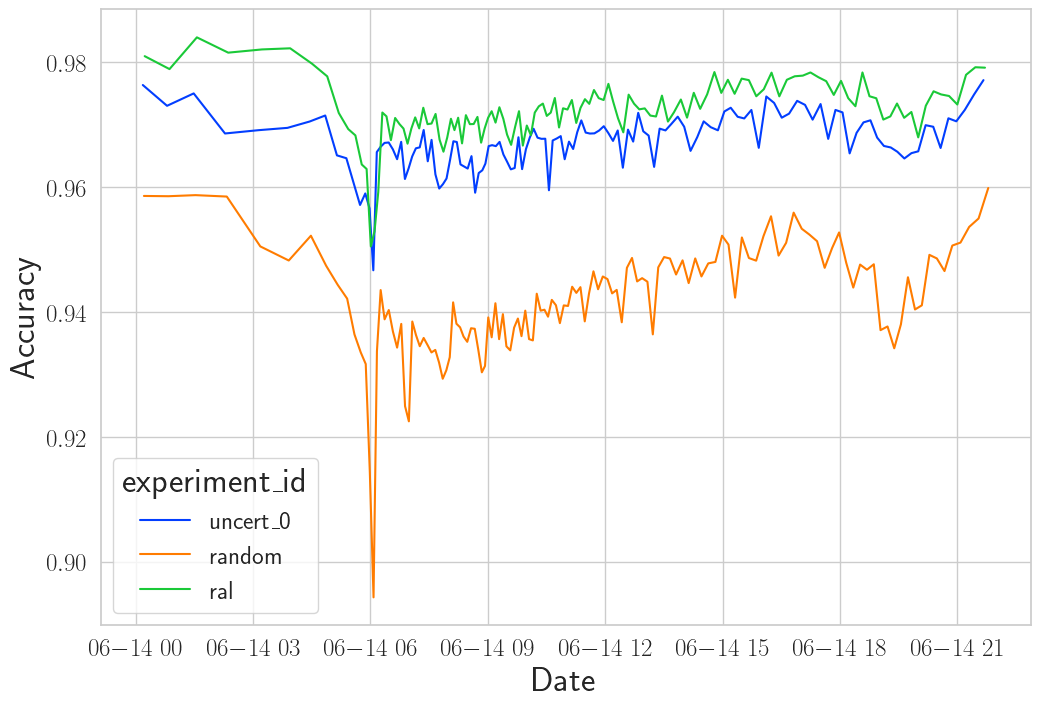}
    \caption{Measurement during one day. Random strategy is obviously weaker than uncertainty based strategies.}
    \label{fig:doh-one-day}
\end{figure}

\begin{figure}
    \centering
        \includegraphics[width=0.45\textwidth]{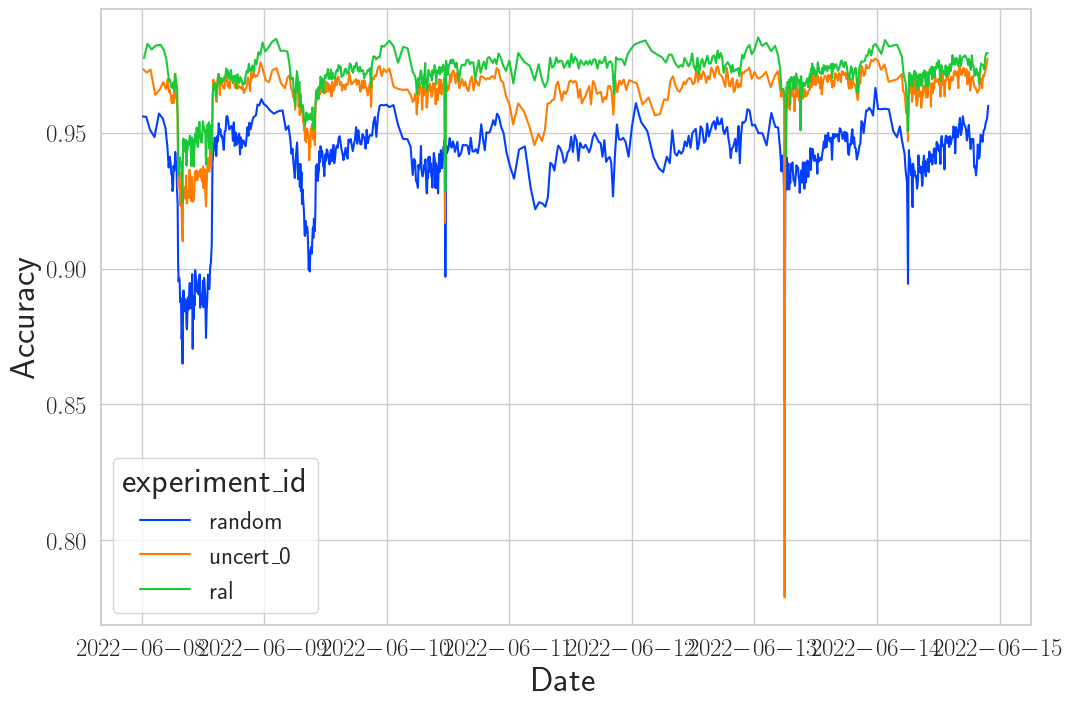}
    \caption{Measurement during one week. Random strategy is not only weaker but also more sensitive -- drops are more deep than uncertainty based strategies.}
    \label{fig:doh-one-week}
\end{figure}

Every day, we observe approximately 3 millions of DoH and 817 millions of TLS flows. Due to the high imbalance ratio, we applied sampling of TLS traffic to be equal to DoH flows. For AL loop we used buffer of 50k records that triggers the loop and selects 10 instances based on the specific strategy. Preview of the obtained results is in Fig. \ref{fig:doh-one-day} as a record of one day; and Fig. \ref{fig:doh-one-week} as a record of one week. We also use dashboard in Grafana for real-time monitoring. We achieved results consistent to the tests in laboratory environment described in Sec.~\ref{sec:improvement}. Random strategy does not work optimally and includes a lot of drops. Contrary, the Uncertainty and RAL strategies are very stable and provides solid results even in a week time frame. It is worth nothing that we do not use any optimization or instance selection method in this use case.

%% file: future.tex
Enhancement of ML classifiers via dataset improvements is very demanding area that needs additional research, however, our first experiments using ALF indicate that it is a promising approach orthogonal to the traditional models and hyperparameters tuning. The deployed ALF under comprehensive continuous monitoring (and visualizing the monitored statistics) revealed several interesting observations that helped to identify the following open research challenges:
\begin{itemize}
    \item \textbf{Continuous Evaluation of Datasets from Real Traffic} ALF is ready to process a stream of flow records, update datasets and machine learning for theoretically infinite time. This way datasets might grow to impractically large sizes that are not feasible for training. Therefore, ALF is ready for the future research of evaluation more advanced strategies and methods that can be automatically applied on datasets to asses their quality.
    \item \textbf{Detection of obsoleted state of the trained machine learning model} Another research potential lies in the detection of state when the machine learning model must be retrained due to some data drift or distribution drift in the network traffic. This area is useful to avoid unnecessary computation for model training that is no better than the current one. On the other hand, timely detection of an obsoleted machine learning model can improve overall performance of the deployed security system containing the machine learning model. In ALF, we have valuable data regarding ML and dataset performance. However, more advanced detection is missing. 
    \item \textbf{Research in datasets optimization and dataset merging} Replacement strategy is related to the optimization of the dataset. Sometimes, there are large datasets that contain too many redundant or similar samples. The larger the dataset, the more time consuming is the training of machine learning models. Therefore, it is beneficial to remove redundant information and decrease size of the dataset. Unfortunately, the removal must be performed carefully to avoid information loss that would prevent training successful machine learning models.
\end{itemize}

%% file: main.bbl
\begin{thebibliography}{10}
\providecommand{\url}[1]{#1}
\csname url@samestyle\endcsname
\providecommand{\newblock}{\relax}
\providecommand{\bibinfo}[2]{#2}
\providecommand{\BIBentrySTDinterwordspacing}{\spaceskip=0pt\relax}
\providecommand{\BIBentryALTinterwordstretchfactor}{4}
\providecommand{\BIBentryALTinterwordspacing}{\spaceskip=\fontdimen2\font plus
\BIBentryALTinterwordstretchfactor\fontdimen3\font minus
  \fontdimen4\font\relax}
\providecommand{\BIBforeignlanguage}[2]{{%
\expandafter\ifx\csname l@#1\endcsname\relax
\typeout{** WARNING: IEEEtran.bst: No hyphenation pattern has been}%
\typeout{** loaded for the language `#1'. Using the pattern for}%
\typeout{** the default language instead.}%
\else
\language=\csname l@#1\endcsname
\fi
#2}}
\providecommand{\BIBdecl}{\relax}
\BIBdecl

\bibitem{NEMEA}
T.~Cejka, V.~Bartos, M.~Svepes, Z.~Rosa, and H.~Kubatova, ``Nemea: A framework
  for network traffic analysis,'' in \emph{2016 12th International Conference
  on Network and Service Management (CNSM)}, 2016, pp. 195--201.

\bibitem{ETA}
B.~Anderson, S.~Paul, and D.~McGrew, ``Deciphering malware's use of tls
  (without decryption),'' \emph{Journal of Computer Virology and Hacking
  Techniques}, vol.~14, 08 2018.

\bibitem{eta-ml}
Z.~Tropková, K.~Hynek, and T.~Čejka, ``Novel https classifier driven by
  packet bursts, flows, and machine learning,'' in \emph{2021 17th
  International Conference on Network and Service Management (CNSM)}, 2021, pp.
  345--349.

\bibitem{Brabec2020}
J.~Brabec, T.~Kom{\'a}rek, V.~Franc, and L.~Machlica, ``On model evaluation
  under non-constant class imbalance,'' \emph{Computational Science – ICCS
  2020}, vol. 12140, pp. 74 -- 87, 2020.

\bibitem{Soukup2021}
D.~Soukup, P.~Tisovčík, K.~Hynek, and T.~Čejka, ``Towards evaluating quality
  of datasets for network traffic domain,'' in \emph{2021 17th International
  Conference on Network and Service Management (CNSM)}, 2021, pp. 264--268.

\bibitem{Settles2009ActiveL}
B.~Settles, ``Active learning literature survey,'' 2009.

\bibitem{AlStudy}
A.~Shahraki, M.~Abbasi, A.~Taherkordi, and A.~D. Jurcut, ``Active learning for
  network traffic classification: A technical study,'' \emph{IEEE Transactions
  on Cognitive Communications and Networking}, vol.~8, no.~1, pp. 422--439,
  2022.

\bibitem{datadrift}
J.~Lu, A.~Liu, F.~Dong, F.~Gu, J.~Gama, and G.~Zhang, ``Learning under concept
  drift: A review,'' \emph{IEEE Transactions on Knowledge and Data
  Engineering}, vol.~31, no.~12, pp. 2346--2363, 2019.

\bibitem{wassermannRALImprovingStreamBased2019}
\BIBentryALTinterwordspacing
S.~Wassermann, T.~Cuvelier, and P.~Casas, ``{{RAL}} - {{Improving Stream-Based
  Active Learning}} by {{Reinforcement Learning}},'' in \emph{European
  {{Conference}} on {{Machine Learning}} and {{Principles}} and {{Practice}} of
  {{Knowledge Discovery}} in {{Databases}} ({{ECML-PKDD}}) {{Workshop}} on
  {{Interactive Adaptive Learning}} ({{IAL}})}. [Online]. Available:
  \url{https://hal.archives-ouvertes.fr/hal-02265426}
\BIBentrySTDinterwordspacing

\bibitem{cardosoRankedBatchmodeActive2017}
\BIBentryALTinterwordspacing
T.~N. Cardoso, R.~M. Silva, S.~Canuto, M.~M. Moro, and M.~A. Gonçalves,
  ``Ranked batch-mode active learning,'' vol. 379, pp. 313--337. [Online].
  Available:
  \url{https://linkinghub.elsevier.com/retrieve/pii/S0020025516313949}
\BIBentrySTDinterwordspacing

\bibitem{Camacho2022}
J.~Camacho and K.~Wasielewska, ``Dataset quality assessment in autonomous
  networks with permutation testing,'' in \emph{IFIP/IEEE International
  Symposium on Integrated Network Management (IM)}, 2022.

\bibitem{Li}
Y.~Li and X.~Chao, ``Distance-entropy: An effective indicator for selecting
  informative data,'' \emph{Frontiers in Plant Science}, vol.~12, 01 2022.

\bibitem{DVRL}
\BIBentryALTinterwordspacing
J.~Yoon, S.~Arik, and T.~Pfister, ``Data valuation using reinforcement
  learning,'' in \emph{Proceedings of the 37th International Conference on
  Machine Learning}, ser. Proceedings of Machine Learning Research, H.~D. III
  and A.~Singh, Eds., vol. 119.\hskip 1em plus 0.5em minus 0.4em\relax PMLR,
  13--18 Jul 2020, pp. 10\,842--10\,851. [Online]. Available:
  \url{https://proceedings.mlr.press/v119/yoon20a.html}
\BIBentrySTDinterwordspacing

\bibitem{Sahu2020}
A.~Sahu, Z.~Mao, K.~Davis, and A.~E. Goulart, ``Data processing and model
  selection for machine learning-based network intrusion detection,'' in
  \emph{2020 IEEE International Workshop Technical Committee on Communications
  Quality and Reliability (CQR)}, 2020, pp. 1--6.

\bibitem{Dudjak2021}
M.~Dudjak and G.~Martinovic, ``An empirical study of data intrinsic
  characteristics that make learning from imbalanced data difficult,''
  \emph{Expert Systems Applications}, vol. 182, p. 115297, 2021.

\bibitem{Cordeiro2020}
F.~R. Cordeiro and G.~Carneiro, ``A survey on deep learning with noisy labels:
  How to train your model when you cannot trust on the annotations?'' in
  \emph{33rd SIBGRAPI Conference on Graphics, Patterns and Images (SIBGRAPI)},
  2020, pp. 9--16.

\bibitem{Gupta2019}
\BIBentryALTinterwordspacing
S.~Gupta and A.~Gupta, ``Dealing with noise problem in machine learning
  data-sets: A systematic review,'' \emph{Procedia Computer Science}, vol. 161,
  pp. 466--474, 2019, the Fifth Information Systems International Conference,
  23-24 July 2019, Surabaya, Indonesia. [Online]. Available:
  \url{https://www.sciencedirect.com/science/article/pii/S1877050919318575}
\BIBentrySTDinterwordspacing

\bibitem{instanceSelection}
J.~Olvera-López, J.~Carrasco-Ochoa, J.~F. Martínez-Trinidad, and J.~Kittler,
  ``A review of instance selection methods,'' \emph{Artif. Intell. Rev.},
  vol.~34, pp. 133--143, 08 2010.

\bibitem{Sun2017}
C.~Sun, A.~Shrivastava, S.~Singh, and A.~Gupta, ``Revisiting unreasonable
  effectiveness of data in deep learning era,'' in \emph{IEEE International
  Conference on Computer Vision (ICCV)}, 2017, pp. 843--852.

\bibitem{muller}
\BIBentryALTinterwordspacing
M.~Muller, I.~Lange, D.~Wang, D.~Piorkowski, J.~Tsay, Q.~V. Liao, C.~Dugan, and
  T.~Erickson, ``How data science workers work with data: Discovery, capture,
  curation, design, creation,'' in \emph{Proceedings of the 2019 CHI Conference
  on Human Factors in Computing Systems}, ser. CHI '19.\hskip 1em plus 0.5em
  minus 0.4em\relax New York, NY, USA: Association for Computing Machinery,
  2019, p. 1–15. [Online]. Available:
  \url{https://doi.org/10.1145/3290605.3300356}
\BIBentrySTDinterwordspacing

\bibitem{MLmonitoring}
\BIBentryALTinterwordspacing
T.~Ginart, M.~Jinye~Zhang, and J.~Zou, ``Mldemon:deployment monitoring for
  machine learning systems,'' in \emph{Proceedings of The 25th International
  Conference on Artificial Intelligence and Statistics}, ser. Proceedings of
  Machine Learning Research, G.~Camps-Valls, F.~J.~R. Ruiz, and I.~Valera,
  Eds., vol. 151.\hskip 1em plus 0.5em minus 0.4em\relax PMLR, 28--30 Mar 2022,
  pp. 3962--3997. [Online]. Available:
  \url{https://proceedings.mlr.press/v151/ginart22a.html}
\BIBentrySTDinterwordspacing

\end{thebibliography}
